\RequirePackage{amsmath} % this first to avoid vec problems
\documentclass[runningheads]{llncs}
\usepackage{graphicx}
\usepackage{booktabs}

\newcommand\T{\rule{0pt}{3.0ex}}       % Top strut
\newcommand\B{\rule[-1.2ex]{0pt}{0pt}} % Bottom strut

\begin{document}
\title{DeepAAA: clinically applicable and generalizable detection of abdominal aortic aneurysm using deep learning}
\titlerunning{DeepAAA (Arxiv version, accepted MICCAI 2019)}

\author{
Jen-Tang Lu\inst{5} \and
Rupert Brooks\inst{4}\orcidID{0000-0002-5642-7770} \and
Stefan Hahn\inst{4} \and
Jin Chen\inst{6} \and
Varun Buch\inst{1,7}\orcidID{0000-0003-1704-4811} \and
Gopal Kotecha\inst{5} \and
Katherine P. Andriole\inst{1,3} \and
Brian Ghoshhajra\inst{2} \and
Joel Pinto\inst{4} \and
Paul Vozila\inst{4} \and
Mark Michalski\inst{1} \and
Neil A. Tenenholtz\inst{1}\orcidID{0000-0003-1250-3716}
}

\authorrunning{J-T. Lu et al. (Arxiv version, accepted MICCAI 2019)}

\institute{
MGH and BWH Center for Clinical Data Science \and
Massachusetts General Hospital (MGH) \and
Brigham and Women's Hospital (BWH) \and
Nuance Communications Inc.
\and work done while affiliated with \textsuperscript{1}
\and work done while affiliated with \textsuperscript{4}
\and corresponding author \email{varun.buch@mgh.harvard.edu}\\
}

\maketitle              % typeset the header of the contribution
\begin{abstract}
We propose a deep learning-based technique for detection and quantification of abdominal aortic aneurysms (AAAs). The condition, which leads to more than 10,000 deaths per year in the United States, is asymptomatic, often detected incidentally, and often missed by radiologists. Our model architecture is a modified 3D U-Net combined with ellipse fitting that performs aorta segmentation and AAA detection. The study uses 321 abdominal-pelvic CT examinations performed by Massachusetts General Hospital Department of Radiology for training and validation. The model is then further tested for generalizability on a separate set of 57 examinations with differing patient demographics and acquisition characteristics than the original dataset. DeepAAA achieves high performance on both sets of data (sensitivity/specificity 0.91/0.95 and 0.85 / 1.0 respectively), on contrast and non-contrast CT scans and works with image volumes with varying numbers of images. We find that DeepAAA exceeds literature-reported performance of radiologists on incidental AAA detection. It is expected that the model can serve as an effective background detector in routine CT examinations to prevent incidental AAAs from being missed.
\end{abstract}
\section{Introduction}
Abdominal aortic aneurysms (AAAs), an enlargement or widening of the abdominal aorta, commonly occurs in males older than 65 years with a prevalence of 4 to 8 percent \cite{lindholtScreeningAbdominalAortic2005}. Untreated aneurysms tend to grow and eventually may rupture with mortality rates exceeding 90$\%$. As most AAAs are asymptomatic until critical bleeding, incidental finding of AAAs becomes critical. However, on routine abdominal computed tomography (CT) exams, only 65$\%$ of AAAs are incidentally identified \cite{claridgeMeasuringAbdominalAortic2017}. This low reporting rate makes it difficult to provide timely intervention for patients. Indeed, it is common for AAAs to be first diagnosed at a point where a patient is already at risk for rupture \cite{mellLateDiagnosisAbdominal2013}. Furthermore, in routine clinical practice, the size of AAAs is determined by manual measurement of the maximal aortic diameter, which is time-consuming and prone to  high inter-reader variability.  

Consequently, a variety of computer-aided diagnosis techniques have been proposed over the past decade for automated aorta segmentation. Many of these previous aids used classical computer vision techniques that required prior knowledge, such as external seed points for initialization \cite{debruijneInteractiveSegmentationAbdominal2004}. Driven by the ever-increasing capability of deep learning, neural networks have recently been used for aorta segmentation on CT angiography \cite{lopez-linaresFullyAutomaticDetection2018}. However, these previous deep learning algorithms focused only on CT exams with contrast, while incidental identification of AAAs on scans without contrast is equally important but more challenging. Additionally, most of the previous works concentrated on the task of automated aortic segmentation \cite{zhugeAbdominalAorticAneurysm2006,siriapisithOuterWallSegmentation2018,lopez-linaresFullyAutomaticDetection2018}, but there are very few studies investigating the more applied task of AAA detection, which has much greater clinical relevance than purely performing segmentation alone.

In this paper, we demonstrate a deep-learning solution (DeepAAA) for automated aorta segmentation and AAA detection on both contrast and non-contrast CT series. Specifically, we develop a variant of a 3D U-Net \cite{cicek3DUNetLearning2016a} for aorta segmentation on abdominal CT scans. The proposed method handles series with varying numbers of images. We then apply ellipse fitting to the segmented aortic contours and estimate the largest aortic diameter. DeepAAA is a general solution, achieving a high detection rate for AAAs on both contrast and non-contrast CT scans and working with variable image resolutions and slice thicknesses. Furthermore, our solution demonstrates strong generalizability and performance relative to literature-reported values for radiologist sensitivity at AAA detection.  

\section{Cohort and annotation}

Image data consisted of contrast and non-contrast CT examinations of the abdomen and pelvis performed between January 2005 and April 2017 by Massachusetts General Hospital Department of Radiology. The investigators obtained local Institutional Review Board approval for the project and selected two datasets from the database. The two datasets differ in terms of their capture dates and imaging equipment used as characterized in Table \ref{tab:datasets}.

\begin{table}[htb]
\caption{Comparison between primary and additional validation data sets \label{tab:datasets}}
\begin{tabular}{lll}
\hline
\textbf{Characteristics:}               & \textbf{Primary Data Set}                                                                  & \textbf{Additional Validation Set}                                                                                            \\ \hline
Number of studies       & 321                                                                                        & 57     \T                                                                                                                       \\
Dates captured          & 2005-2007 (90\%)                                                                           & 2012-2016 (85\%)   \T  \B                                                                                                         \\
%                        &                                                                                            &                                                                                                                               \\
\begin{tabular}[c]{@{}l@{}}Imaging equipment \\
manufacturer\end{tabular}\T& \begin{tabular}[c]{@{}ll@{}} A 96\%&  B 4\%\end{tabular}           & \begin{tabular}[c]{@{}ll@{}} A 61\% &  B 26\% \\  C 8\% & D 5\%\end{tabular} \\
%                        &                                                                                            &                                                                                                                               \\
Contrast \%             & 48\%                                                                                       & 51\%                                                                                                                       \T   \\
%                        &                                                                                            &                                                                                                                               \\
Presence of AAA \%      & 77\%                                                                                       & 51\%                          \T                                                                                                \\
Mean age (by study)        & 70 years                                                                                   & 72 years        \T                                                                                                              \\
%                        &                                                                                            &                                                                                                                               \\
Gender (by study)          & 68\% Male, 32\% Female                                                                     & 68\% Male, 32\% Female      \T   \B                                                                                              \\
%                        &                                                                                            &                                                                                                                               \\
\hline
\textbf{Data labelling method:}  &                                                            &                                                                                                                               \\ \hline

Max. aortic diameter & Manual segmentation                                                                        & Sourced from reports  \T  \B                                                                                                      \\ 
\begin{tabular}[c]{@{}l@{}}Presence/absence\\ of AAA\end{tabular} \T& \begin{tabular}[c]{@{}l@{}}3.0 cm threshold applied~~\\ to segmentation\end{tabular}                                                    & Sourced from reports                                                                                                          \\ \hline

\end{tabular}
\end{table}

\subsection{Primary Data Set}

The primary dataset was used for the training and initial validation of the model and contained  321 studies (223 unique patients). These were selected based on a keyword search of study reports ensuring a mixture of positive and negative cases of AAA. The query was biased to largely include studies captured between 2005 and 2007. Of the studies selected, there were 217 (67.6 $\%$) males and 104 (32.4 $\%$) females with a mean age of 70.3 years; 153 (47.7 $\%$) CT scans with contrast and 168 (52.3 $\%$) without; 247 (76.9 $\%$) studies with AAA present and 74 (23.1 $\%$) without AAA. For each study, the axial series was used for aorta segmentation and AAA detection. Slice thickness of the images ranged from 2 to 10 mm, while the number of images for each series varied from 40 to 384.

To generate a ground-truth aortic segmentation, the abdominal aorta was manually contoured on the axial scans slice-by-slice until the aortic bifurcation. Each study was annotated by 1 to 4 CT technologists under supervision of 2 radiologists. Based on the clinical definition \cite{claridgeMeasuringAbdominalAortic2017}, the presence of AAA was determined by applying a 3.0 cm threshold to the maximum aortic diameter as defined by the manual segmentations. 

As many exams were annotated by multiple annotators, a partial assessment of inter-rater variability was possible.  Of the 153 contrast studies, 124 were annotated by at least 2 independent technologists, leading to 517 pairwise comparisons.  The non-contrast data, however, contained only 10 studies where more than one segmentation was performed, resulting in only 16 pairwise comparisons.  The average inter-rater Dice on contrast series was $0.95\pm0.03$, while on noncontrast series, it was $0.90\pm0.08$.  Given the small number of samples, the inter-rater variability on non-contrast data should not be considered definitive but suggests roughly similar levels of agreement. For the subsequent analysis, one reference segmentation per dataset was selected randomly as ground truth.

\subsection{Additional Validation Set}

An additional validation set was used to test the robustness of the model to changes in imaging equipment, imaging department capture protocols, and patient demographics. All of these factors may vary significantly over time at a single site, and thus, we selected 57 studies (57 unique patients) predominantly captured between 2012 and 2016 for this dataset. The studies were selected to include a mixture of positive and negative cases of AAA through keyword search of study reports. All negative studies were manually verified to not contain a AAA. To assess the model against radiologist-reported ground truth and validate post-processing stages which generate the AAA measurement, the maximum aortic diameter and presence of AAA was sourced from radiology reporting rather than being derived from manual segmentations (as was done for the primary data set).

\section{Methods}

We achieve AAA detection via two sequential steps:  (1) aorta segmentation (2) aorta contour fitting for the estimation of the largest cross-sectional diameter. For abdominal aortic segmentation, we developed a variant of a 3D U-Net \cite{cicek3DUNetLearning2016a} which accepts series with varying numbers of images. As discussed in Section 2, our dataset contained a wide distribution of image counts and slice thicknesses as abdominal studies may also cover other regions of the body, including the pelvis or thorax. It is thus essential to develop an algorithm adapts to variability along the axial dimension. The 3D U-Net architecture we used contained 4 down/upsampling modules (plus the bottleneck layer), 2 convolutional layers per module, and 32 initial features in the network. The convolutional kernel size was 3 $\times$ 3 $\times$ 3 in both the downsampling and upsampling path, while the 3D pooling kernels were 2 $\times$ 2 $\times$ 1 to preserve image count. Batch normalization was applied before each ReLU activation, and dropout regularization was utilized at the bottleneck layer with a dropout rate of 0.2. A 1 $\times$ 1 $\times$ 1 convolutional layer with softmax activation over two classes (background and aorta) was applied at the output layer and thresholded at 0.5 to generate the binary aorta mask. 

The model was trained with the RMSprop optimizer using a learning rate of 0.0001.   Weights selected for evaluation were those that minimized the loss on the validation set, which were not in general the last epoch weights.  The loss function was a smoothed negative Dice coefficient:
\begin{equation}
D=-\frac{2 \sum_{i=1}^N p_i g_i + 1}{\sum_{i=1}^N p_i + \sum_{i=1}^N g_i + 1}
\end{equation}
similar, but not identical, to that used in \cite{milletariVNetFullyConvolutional2016b}.  The summation is over all $N$ voxels in a scan, $p_i$ is the predicted aorta probability and $g_i$ is the ground truth classification for voxel $i$. The additional ones in the numerator and denominator avoid division by zero and yield a perfect score for a correct, empty segmentation. 

In order to build a general AAA detector that worked with both contrast and non-contrast CT scans, we mixed both types of CT images for model training. All the experiments were implemented utilizing the Keras deep learning library with the Tensorflow backend on NVIDIA DGX-1 Volta.

After aorta segmentation, we applied ellipse fitting \cite{fitzgibbonDirectLeastSquares1996} image-by-image to the contours of the aorta. The largest aortic diameters (d) were thus assigned by the long axis of the ellipses. For the regions where the aorta was not parallel to the axial CT scans, angle correction was applied to retrieve the true aorta diameter, i.e. $d \cos \theta$, where $\theta$ was the angle between the secant plane of the aorta and the axial scan. Based on the definition of AAA, predicted positives were the studies where the largest diameter of the aorta segment was greater than 3cm. We then compared the predicted results with the ground truth annotations.

\section{Results}

\subsection{Training and Cross-Validation on Primary Data Set}
To assess model validity and repeatability, the primary dataset was divided into 5 folds such that no patient was repeated between folds. Cross validation was performed by selecting folds $\{n,n+1,n+2\} \bmod 5$ as training, $n+3 \bmod 5$ as validation and the remaining fold as test for $n\in \{0..5\}$.  For each combination, the weights with the best validation score after 100 epochs were selected.

\begin{table}[tb]
  \caption[Results of 5-fold crossvalidation]
  {Results of 5-fold cross-validation. Delta is predicted minus reference largest diameter.  Standard deviations combined using pooled variance.  
} 
  
  \label{tab:crossvalidation}
  \centering
  \begin{tabular}{c@{\hskip 0.25in}c@{\hskip 0.25in}c@{\hskip 0.25in}c}
    \toprule
    Fold & N & Mean Dice & Mean Delta (mm) \\
    \midrule
    0&64 & 0.887 $\pm$ 0.121 & -0.4 $\pm$ 8.5 \\
    1&64 & 0.893 $\pm$ 0.107 & -0.7 $\pm$ 5.4 \\
    2&64 & 0.894 $\pm$ 0.060 & -3.2 $\pm$ 6.0 \\
    3&64 & 0.883 $\pm$ 0.126 & -2.7 $\pm$ 6.3 \\
    4&63 & 0.877 $\pm$ 0.127 & 0.8 $\pm$ 9.5 \\
    \midrule
    All\footnotemark &319 & 0.887 $\pm$ 0.111    & -1.3 $\pm$ 7.3 \\
  \bottomrule
  \end{tabular}
\end{table}
\footnotetext{Total is not 321 as two datasets were excluded due to truncated images.  They were retained in the generation of the full model.}

Inference on each test study was evaluated in terms of Dice score relative to the reference segmentation and in terms of the maximum diameter of the aorta evaluated on the inferred segmentation versus the same calculation on the reference segmentation.
The detailed results of this cross validation are presented in Table~\ref{tab:crossvalidation}.  Over the 5 folds, the average Dice score ranged from 0.883 to 0.894, with a average Dice score of $0.887\pm0.111$.  The estimate of the diameter is consistently within one standard deviation of zero.  There may be a slight bias towards smaller diameter, as 4 of the 5 folds had negative means but this bias is small with overall mean -1.3mm $\pm$ 7.3.  

For a final set of weights, the complete primary dataset was randomly split into training (80$\%$), validation (10$\%$), and test sets (10$\%$). Training was performed for 300 epochs and the weights with lowest validation loss were selected.

\begin{figure}[!t]
  \centering 
  \includegraphics[width=4.5in]{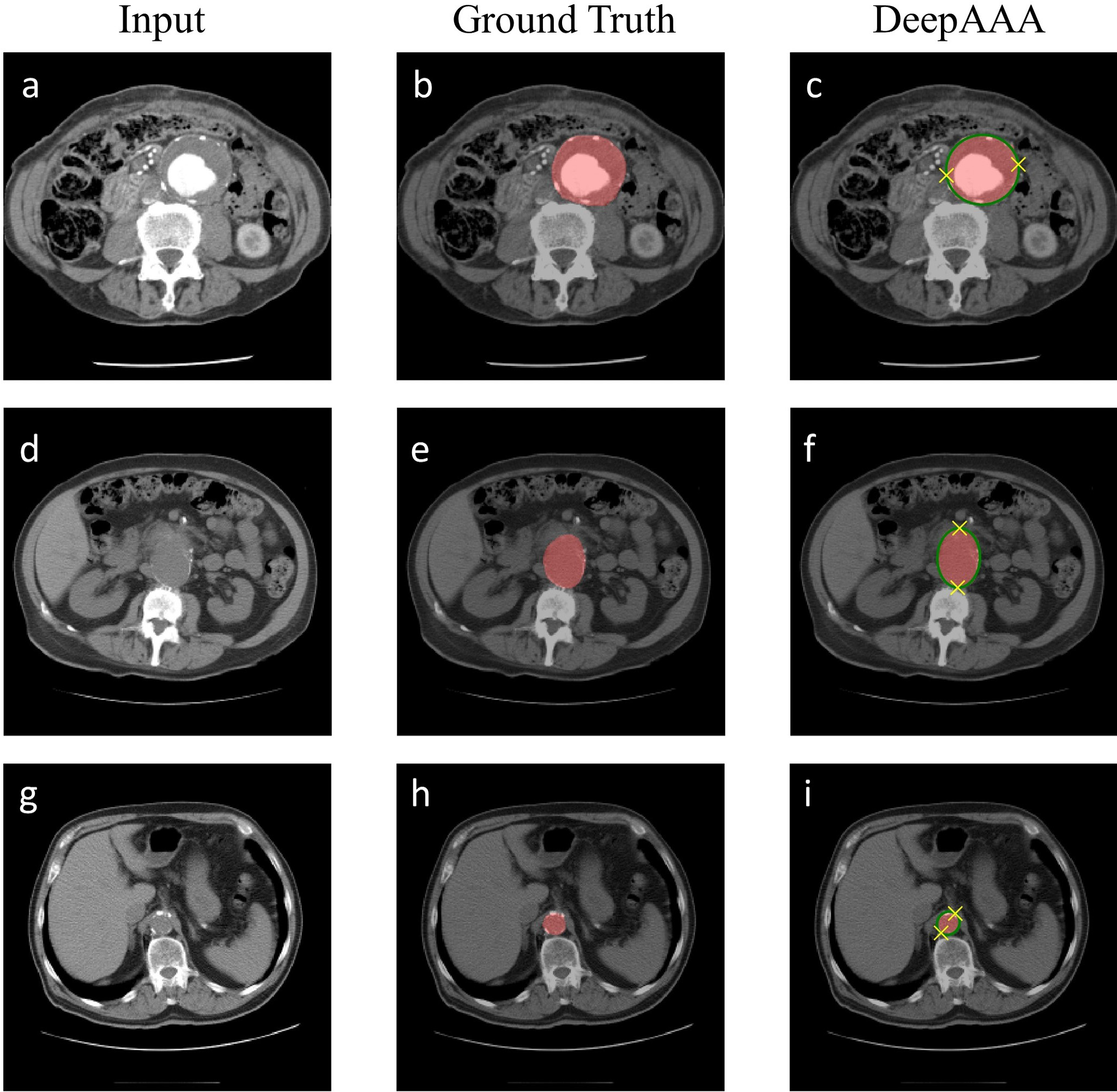} 
  \caption{DeepAAA aorta segmentation (red overlay) and the largest aortic diameter estimation (yellow crosses, the long axis of ellipse fitting [green curves] of the aorta segment): (a-c) Aneurysm with thrombus on contrast CT. (d-f) Large aneurysm on non-contrast CT where aortic boundary is hard to segment. (g-i) normal aorta.}
  \label{fig:results} 
\end{figure}

As shown in Fig. 1, DeepAAA successfully segments the aorta on both contrast and non-contrast CT images, and works well with more challenging cases where blood-clots are present or the aortic boundary is unclear in the images. We achieve high performance on aortic segmentation with an average Dice coefficient of 0.91, which yields high sensitivity (0.91) and high specificity (0.95) on AAA detection (Table \ref{tab:performance}). We further examine the error in the largest aortic diameter measurement ($d_{pred}$ – $d_{true}$). We find that the algorithm tends to underestimate the aorta size, but the 2.02 mm average discrepancy is well within the 10 mm gradations on which clinical decisions are generally based.

\begin{table}[!t]
  \caption{Performance of DeepAAA on segmentation and detection}
  \label{tab:performance}
  \centering
  \begin{tabular}{ll@{\hskip 4mm}cc@{\hskip 4mm}cc}
    \toprule
    \multicolumn{2}{c}{} & \multicolumn{2}{c}{Segmentation} & \multicolumn{2}{c}{Detection of AAA}\\
    \cmidrule(r){3-4} \cmidrule(r){5-6}
Dataset &    CT Type     &  Dice    & Mean Delta (mm) & Sensitivity & Specificity\\
    \midrule
{\bf Primary}&    Contrast & 0.89 $\pm$ 0.05 & -2.67 $\pm$ 2.62    & 0.89  & 0.94  \\
    & Non-contrast & 0.90 $\pm$ 0.05 & -1.36 $\pm$ 4.30      & 0.92  & 0.95  \\
    \cmidrule(r){2-6}
    &{\bf Overall}     & {\bf 0.90 $\pm$ 0.05}      & {\bf -2.02 $\pm$ 3.62}  & {\bf 0.91 } & {\bf 0.95 } \\
    \midrule
      \multicolumn{3}{l}{\bf Additional Validation Set}  &{\bf -0.6 $\pm$ 3.0}  & {\bf 0.85} & {\bf 1.00} \\
 \bottomrule
  \end{tabular}
\end{table}

\subsection{Testing Model Robustness on the Additional Validation Set}

Using the final model trained in Section 4.1, we performed inference on studies from the additional validation set described in Section 2.2. Each study was labelled for the presence of a AAA via the radiology report, and for those studies with positive findings, the maximum aortic diameter was also extracted.

For each study, the model's outputs were compared to the study labels and the model's overall performance was measured in terms of sensitivity/specificity for detecting AAA and mean error in the maximum diameter. Table \ref{tab:performance}, last row, summarizes these results, along with a comparison to the model's performance on the held-out test set for the same metrics. During the process we noted that some studies in this additional validation set extended into thoracic anatomy, and model inference of this region was removed manually in post-processing.

\section{Discussion}
While AAAs are rarely missed when the leading indication for a study, the rate of detection significantly decreases when the AAA is an incidental finding. DeepAAA aims to provide a ``second set of eyes'' and reduce the rate of missed incidental findings. Therefore, to properly contextualize model performance, it is important to quantify this rate of misdiagnosis. Claridge et al, in a retrospective analysis of 3246 abdominal CT scans and their reports, found that only 65\% of AAAs were detected by radiologists \cite{claridgeMeasuringAbdominalAortic2017}. DeepAAA exceeds the sensitivity they found (Table \ref{Comparison}) while achieving a high specificity (Table \ref{tab:performance}) and localizes the suspected AAA for radiologist confirmation. Thus, a parallel read from our algorithm could potentially provide a significant reduction in missed AAAs and offer significant clinical value, enabling early detection and treatment of AAA. 

Many observers have noted that machine learning models applied to radiology may not generalize well \cite{zechgeneral2018}. Changing the equipment used to capture input images and changing the demographics of the underlying patient cohorts tend to reduce model performance. This lack of generalizability would significantly hamper a model's clinical utility because deployment at sites other than where the model was trained may result in surprising under-performance. To test DeepAAA's ability to generalize, we simulated a significant change in input data by creating a second cohort of validation data (Section 2.2) acquired from different patients using different equipment more than five years after the original training data were acquired. The model showed higher specificity (100\%) and reduced mean error in diameter prediction with only slightly lower sensitivity (85\%) - essentially demonstrating that the model is robust and has not over-fit to any cohort- or equipment-related idiosyncrasies of the original training data.  
\begin{table}[!t]
  \caption{Comparison between DeepAAA and literature reported performance of radiologists on AAA reporting for routine abdominal CT according to aneurysm size}
  \label{Comparison}
  \centering
  \begin{tabular}{cccc}
    \toprule
    Method     &  30-39 mm    & 40-49 mm & $\geq$50 mm \\
    \midrule
    {\bf DeepAAA sensitivity} & {\bf 0.68}  & {\bf 1.00}     & {\bf 1.00} \\
    Radiologists' sensitivity\cite{claridgeMeasuringAbdominalAortic2017}  & 0.52 & 0.87 & 1.00 \\
    \bottomrule
  \end{tabular}
\end{table}

Future work would involve extending the DeepAAA model beyond the abdominal region to include segmentation of the thoracic aorta. Thoracic aortic aneurysms (TAA), although not nearly as prevalent as AAA, are still a significant source of mortality and generally affect a younger population. In addition, models to predict AAA growth or rupture would be of significant clinical value in guiding more targeted surveillance programs and therapy.

\section{Supplemental Material: Revised Cross Validation results}
In the main paper, the cross validation results presented in Table 2 were slightly inconsistent with 
the remainder of the paper as two datasets were omitted due to differences in processing techniques.  
In this supplement, we present the cross validation results on the full primary dataset to avoid any confusion related to this issue.  While some small numerical changes did occur, the overall conclusions remain the same.

\begin{table}[htb]
  \caption[Results of 5-fold crossvalidation]
  {Results of 5-fold cross-validation. Delta is predicted minus reference largest diameter.  Standard deviations combined using pooled variance.  
} 

  \label{tab:crossvalidationsupp}
  \centering
  \begin{tabular}{c@{\hskip 0.25in}c@{\hskip 0.25in}c@{\hskip 0.25in}c}
    \toprule
    Fold & N & Mean Dice & Mean Delta (mm) \\
    \midrule
    0&65 & 0.869 $\pm$ 0.143 & 0.7 $\pm$ 10.0 \\
    1&64 & 0.848 $\pm$ 0.170 & -2.1 $\pm$  8.9\\
    2&64 & 0.864 $\pm$ 0.155 & -2.7 $\pm$ 10.4 \\
    3&64 & 0.901 $\pm$ 0.059 & -2.4 $\pm$ 5.2 \\
    4&64 & 0.882 $\pm$ 0.078 & -2.0 $\pm$ 8.3 \\
    \midrule
    All &321 & 0.873 $\pm$  0.129   &  -1.7 $\pm$ 8.7 \\
  \bottomrule
  \end{tabular}
\end{table}

The primary dataset was divided into 5 folds such that no patient was repeated between folds.  Note that the folds in this supplement are not the same folds as those in Table 2 in the main paper, the difference was necessary to maintain a balanced number of datasets per fold while also not allowing any patient to be present in more than one fold. Cross validation was performed by selecting folds $\{n,n+1,n+2\} \bmod 5$ as training, $n+3 \bmod 5$ as validation and the remaining fold as test for $n\in \{0..5\}$.  For each combination, the weights with the best validation score after 100 epochs were selected.

Inference on each test study was evaluated in terms of Dice score relative to the reference segmentation and in terms of the maximum diameter of the aorta evaluated on the inferred segmentation versus the same calculation on the reference segmentation.
The detailed results of this cross validation are presented in Table~\ref{tab:crossvalidationsupp}.  Over the 5 folds, the average Dice score ranged from 0.848 to 0.901, with a average Dice score of $0.873\pm0.129$.  The estimate of the diameter is consistently within one standard deviation of zero.  There may be a slight bias towards smaller diameter, as 4 of the 5 folds had negative means but this bias is small with overall mean -1.7mm $\pm$ 8.7.

% References
\bibliographystyle{splncs04}
\bibliography{DeepAAA}

\end{document}